\documentclass[aps,prb,twocolumn,superscriptaddress,showpacs,floatfix,
]{revtex4}
\usepackage{amssymb}
\usepackage{graphicx}
\usepackage{amsmath}
\usepackage{times}
\usepackage{color}
\usepackage{setspace}
\usepackage{bm}
\usepackage{epsfig}
\usepackage{amsmath}
\usepackage[breaklinks=true,colorlinks,citecolor=blue,linkcolor=blue,urlcolor=blue]{hyperref}
\def\be{\begin{equation}}
\def\ee{\end{equation}}

\def\ber{\begin{eqnarray}}
\def\eer{\end{eqnarray}}

\begin{document}
\newcommand {\bqa} {\begin{eqnarray}}
\newcommand {\eqa} {\end{eqnarray}}
\newcommand {\ba} {\ensuremath{b^\dagger}}
\newcommand {\no} {\nonumber}
\newcommand{\kk}{{\bf k}}

\title{Tunable wavevector filtering in borophane based normal metal-barrier-normal metal junctions}

\author{Prasun Das}
\email{dasprasun79@gmail.com}
\affiliation{Department of Physics, Jadavpur University, 188 Raja Subodh Chandra Mallick Road, Kolkata 700032, India.}
 
\author{Sangita De Sarkar}
\email{sangitads1185@gmail.com}
\affiliation{Department of Physics, Asansol Girls' College, Dr. Anjali Roy Sarani, Asansol- 713304, West Bengal, India}

\author{Asim Kumar Ghosh}
\email{asimkumar96@yahoo.com}
\affiliation{Department of Physics, Jadavpur University, 188 Raja Subodh Chandra Mallick Road, Kolkata 700032, India.}

\begin{abstract}
We study the transport properties of Dirac electrons across a two-dimensional normal metal-barrier-normal metal (NBN) interfaces in monolayer borophane. We analyse the transmission probability with variation of the width of the barrier region, the incidence energy and transverse momentum. We demonstrate that a gap exists in the transmission probability spectrum and the position, width of transmission gap can be tuned by the barrier strength and transverse momentum respectively. We point out the variation of the ballistic tunneling conductance as a function of the width of the barrier region and incident energy. We find that the oscillatory or decaying nature of the conductance with variation in barrier width depends upon the number of propagating and evanescent modes which are controlled by the incident energy and barrier strength. We show that the conductance as a function of incident energy drops to a minimum value when the incident energy becomes identical to the barrier height and identify that this effect is caused by the presence of evanescent modes inside the barrier. Based on these findings we propose a perfectly tunable wavevector filter for Borophane. We expect our findings posses useful applications in borophane based nano-electronic devices.

\end{abstract}

\maketitle
\section{Introduction} \label{sec1}

Over the last decade, after the discovery of graphene [\onlinecite{novo}], two-dimensional Dirac materials [\onlinecite{2dmaterial1}--\onlinecite{2dmaterial5}] such as graphene and topological insulators [\onlinecite{topology}] appear to be a subject of intriguing interest in both theoretical and experimental condensed matter physics. The low-energy quasiparticles of such two-dimensional materials behave as relativistic Dirac particles [\onlinecite{diraceqn1}--\onlinecite{diraceqn5}]and henceforth the materials exhibit numerous exotic signatures like unconventional quantum Hall effect  [\onlinecite{qhall1}--\onlinecite{qhall4}], minimum conductivity [\onlinecite{qhall2}], unusual Kondo effect [\onlinecite{kondo1}--\onlinecite{kondo5}], exceptional transport properties [\onlinecite{exceptionaltransport1}--\onlinecite{exceptionaltransport5}] $etc$.

Borophane is one such two-dimensional material, fabricated from boron. Among various nanostructures of boron, a $2$D graphenelike crystal called borophene has been investigated extensively [\onlinecite{borophene1}--\onlinecite{borophene3}] due to its unconventional asymmetry feature compared to graphene. The honeycomb lattice structure of boron is unstable due to electron deficiency. However, stable lattice structure can be obtained with introduction of additional boron atoms. This leads to many possible allotropes such as $\alpha$ sheet, $\beta$ sheet, $\beta_{12}$ sheet, 8-$Pmmn$ borophene $etc$. Signature of anisotropic massless Dirac Fermions have been exhibited in the  $\beta_{12}$ sheet [\onlinecite{betadirac}] and also in the 8-$Pmmn$ borophene [\onlinecite{8pmmndirac}].
The presence of imaginary frequencies in the phononic dispersion of borophene leads to unstability of the material against the periodic vibrations with long wavelength [\onlinecite{imaginaryf1}, \onlinecite{imaginaryf2}]. Surface hydrogenation of borophene is one feasible method to construct stabilised borophene. According to Xu $et. al$.[\onlinecite{imaginaryf1}], fully hydrogenated borophene ($B_2H_2$), called borophane, is a stable structure which is produced in vacuum without adding a substrate to borophene. The behaviour of borophane as Dirac material with a remarkable Fermi velocity, twice to four times of that of graphene is also exhibited [\onlinecite{imaginaryf1}, \onlinecite{velocity1}, \onlinecite{velocity2}]. Density functional theory (DFT) calculations show the existence of an anisotropic tilted Dirac cone in borophane [\onlinecite{imaginaryf1}, \onlinecite{diracconeborophene}]. The anisotropy feature of borophane is also manifested in its mechanical properties [\onlinecite{mechanical1}--\onlinecite{mechanical3}], electrical properties [\onlinecite{electrical1}, \onlinecite{electrical2}], magnetic properties [\onlinecite{magnetic1}--\onlinecite{magnetic3}], optical properties [\onlinecite{charlier},\onlinecite{opt2}] and superconductivity [\onlinecite{superconductivity1}--\onlinecite{superconductivity4}]. Along the valley direction (armchair), the current- voltage characteristic of borophane is linear showing a metallic trend in contrast to the buckled direction, exhibiting semiconductor nature [\onlinecite{semiconductor1}--\onlinecite{semiconductor3}]. 

The presence of such anisotropy along with the tilt in the material with a high Fermi velocity motivates us to study whether the tilt affects the transport properties. The ballistic normal metal-barrier-normal metal (NBN) junction is the basic constituent of various novel devices, in which Dirac Fermions exhibits a lot of unusual features and provides a platform to study the transport properties. A completely different nature between Dirac and Schr\"{o}dinger quasiparicles can be explored using NBN junctions. Although the barrier strength has a higher value than incident energy, the transmission probability in NBN junction and the conductance show oscillatory nature for Dirac like quasiparticles in contrast to the exponentially decaying nature of Schr\"{o}dinger quasiparticles. 

In this work, we have studied the transport properties of Dirac electrons in borophane across a NBN junction. Since the tight binding hamiltonian of borophane possesses anisotropy in momentum space, the orientation of the NBN junction should have an important effect in the transport properties. The presence of the linear $k_x$ term (tilt) in dispersion relation stimulates us to choose the orientation of the junction along the $\Gamma-X$ direction. Our motivation is to explore how the tilted dispersion affects the formalism and hence the transport properties of a NBN junction in borophane. We have found that the presence of tilting term in Hamiltonian has a considerable effect on the formalism of transport. One unique feature to emphasise is that an anisotropy is introduced between the angle of reflection and the angle of incidence. The isotropic case of mirror reflection is modified in this case and the formalism reduces to that of isotropic non-tilted material such as graphene as a limiting case when the tilt term vanishes. We have analytically studied the transmission probability and then the ballistic tunneling conductance is found out numerically. As already stated, in the limit when the tilt term vanishes, our result agrees with that of graphene, as expected. We find that the transmission probability is an oscillatory or a decaying function of the barrier width depending upon the incident energy, barrier strength and the transverse momentum. We have presented a condition of transition between these two phases and the phases are shown in a phase diagram as a function of a critical parameter. The conductance also exhibits the transition from oscillatory to decaying region as a function of barrier width. There exists a transmission gap in the transmission spectrum due to the presence of evanescent modes within the barrier. The appearence of evanescent modes leads to minima in the conductance spectrum. We have investigated the modulation of transmission gap by varying the model parameters like the incident energy, barrier strength and the transverse momentum. We present a discussion on how the nature of conductance depends on the nature and number of contributing transmission modes if one varies incident energy and barrier strength. Our results show that one can effectively design a tunable wavevector filter [\onlinecite{wavefilter1}--\onlinecite{wavefilter6}] using borophane based NBN junction.

The rest of the paper is organised as follows. In section \ref{sec2}, we explain the model, formalism and the analytical calculation of transmission probability starting from low energy effective Hamiltonian of borophane. Then in section \ref{sec3}, we have discussed our analytical and numerical results. Lastly, we conclude and abridge our findings in section \ref{sec4}. 
\begin{figure}[t]
\begin{center}
\includegraphics[width=.95\linewidth]{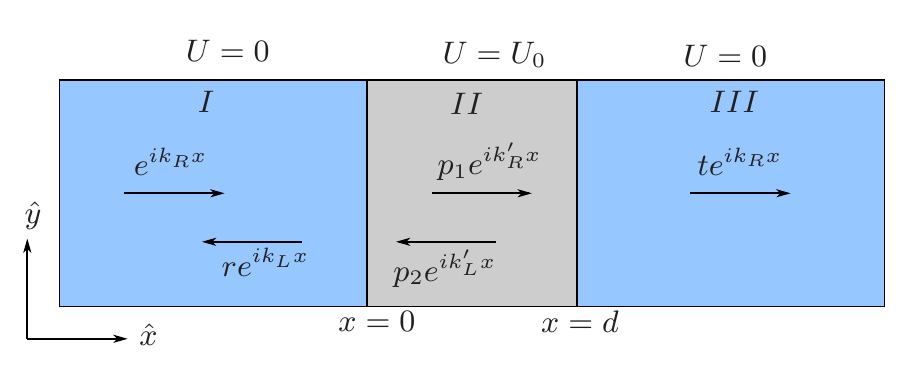}
\caption{Pictorial representation of normal metal-barrier-normal metal junction. Blue regions (region $I$ and $III$) represent normal metal portion of the junction, where there is no external potential barrier $(U=0)$. In the gray region of width $d$  an external potential $U_0$ exists. The incident, reflected and transmitted wave functions are shown in each region. See text for details.
\label{nbn}}
\end{center}
\end{figure}

\section{Model and Formalism} \label{sec2}
\label{form}
\begin{figure}[t]
\begin{center}
\includegraphics[width=.95\linewidth]{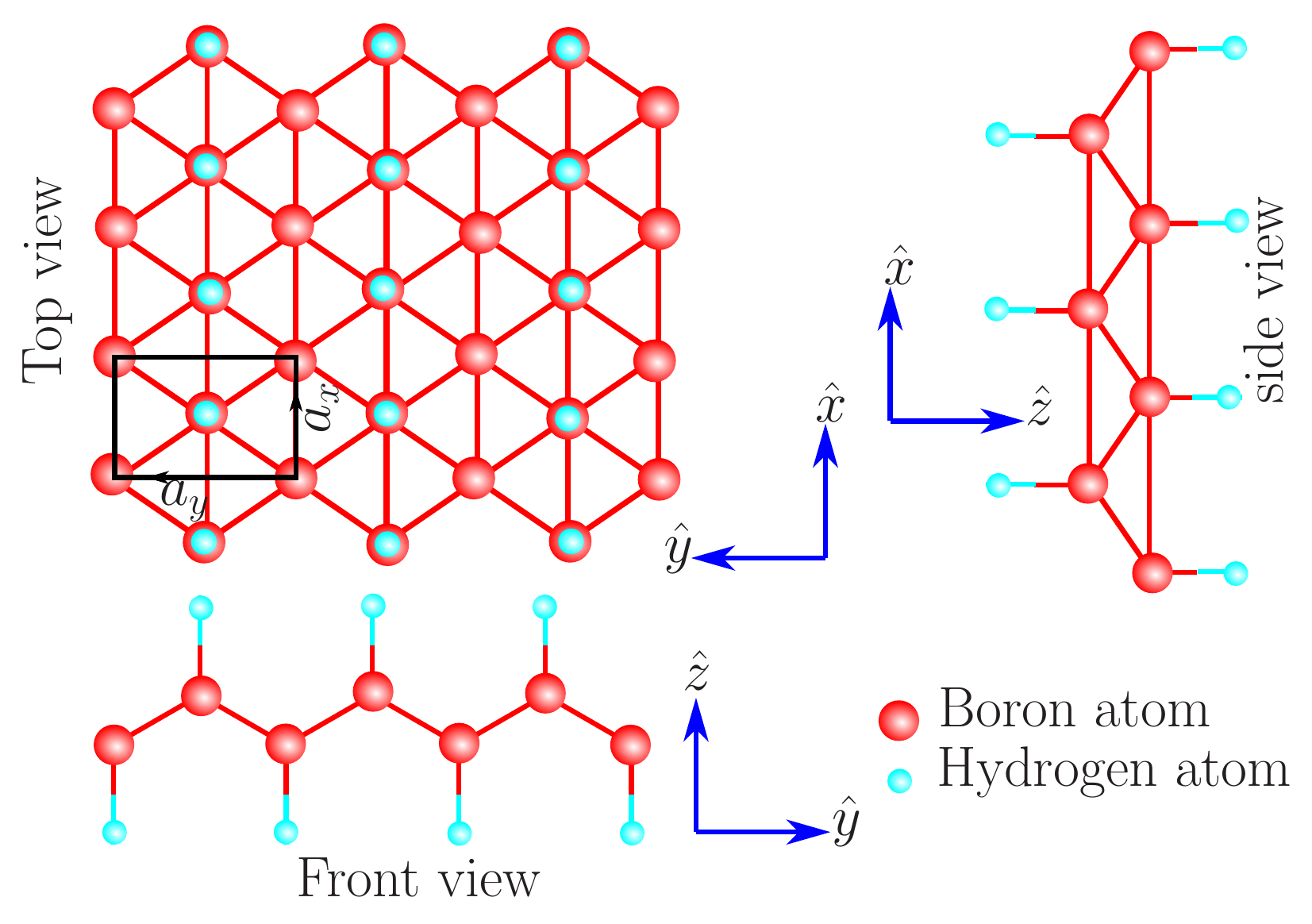}
\caption{Schematic illustration of borophane lattice structure exhibiting top view, side view and front view. The unit cell is shown as the black rectangle which contains two boron atoms (red balls) and two hydrogen atoms (aqua balls). Here $a_x$, $a_y$ represents the primitive lattice constants along $x$ and $y$-directions, respectively.
 \label{unitcell}}
\end{center}
\end{figure}

\begin{figure}[t]
\begin{center}
\includegraphics[width=1.\linewidth]{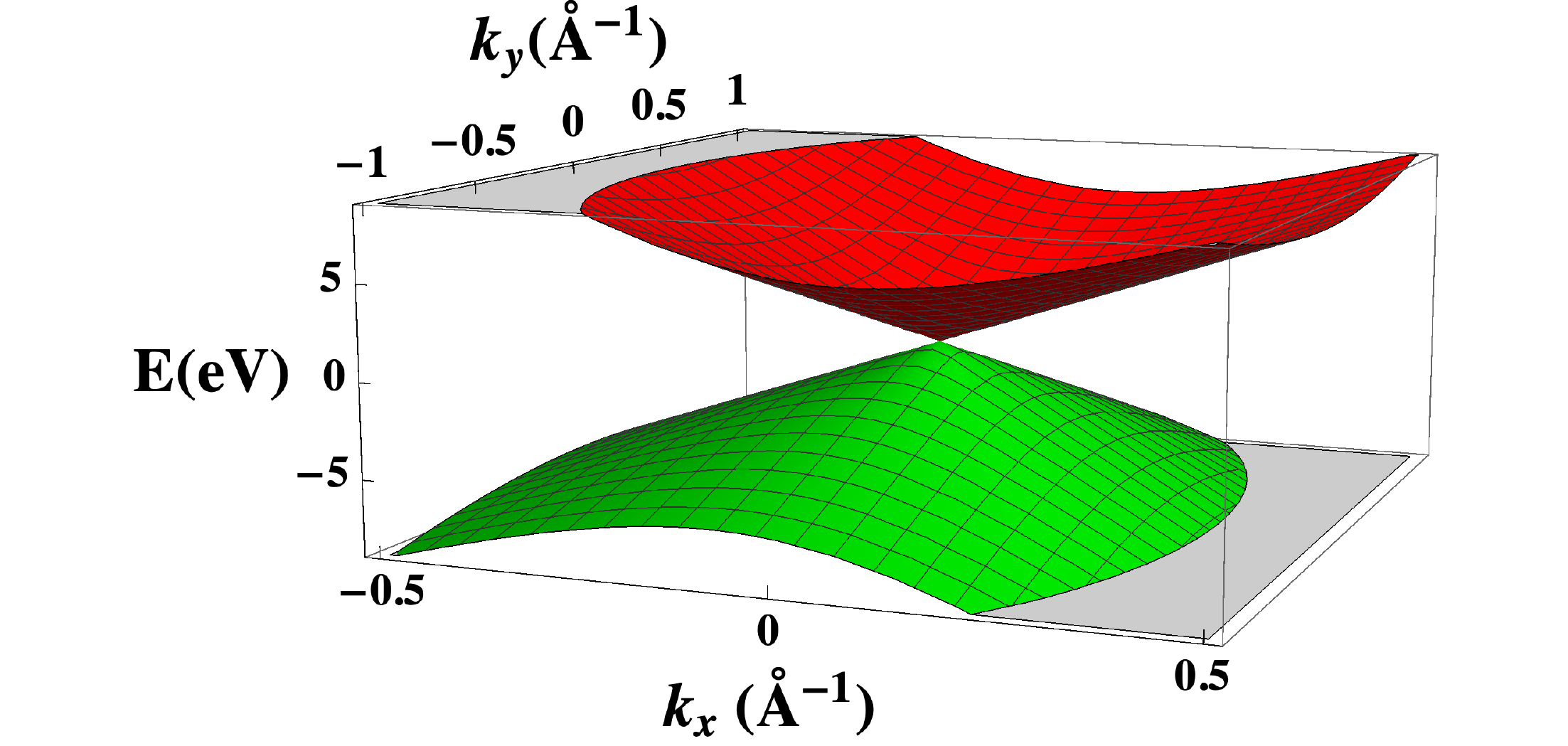}
\includegraphics[width=.95\linewidth]{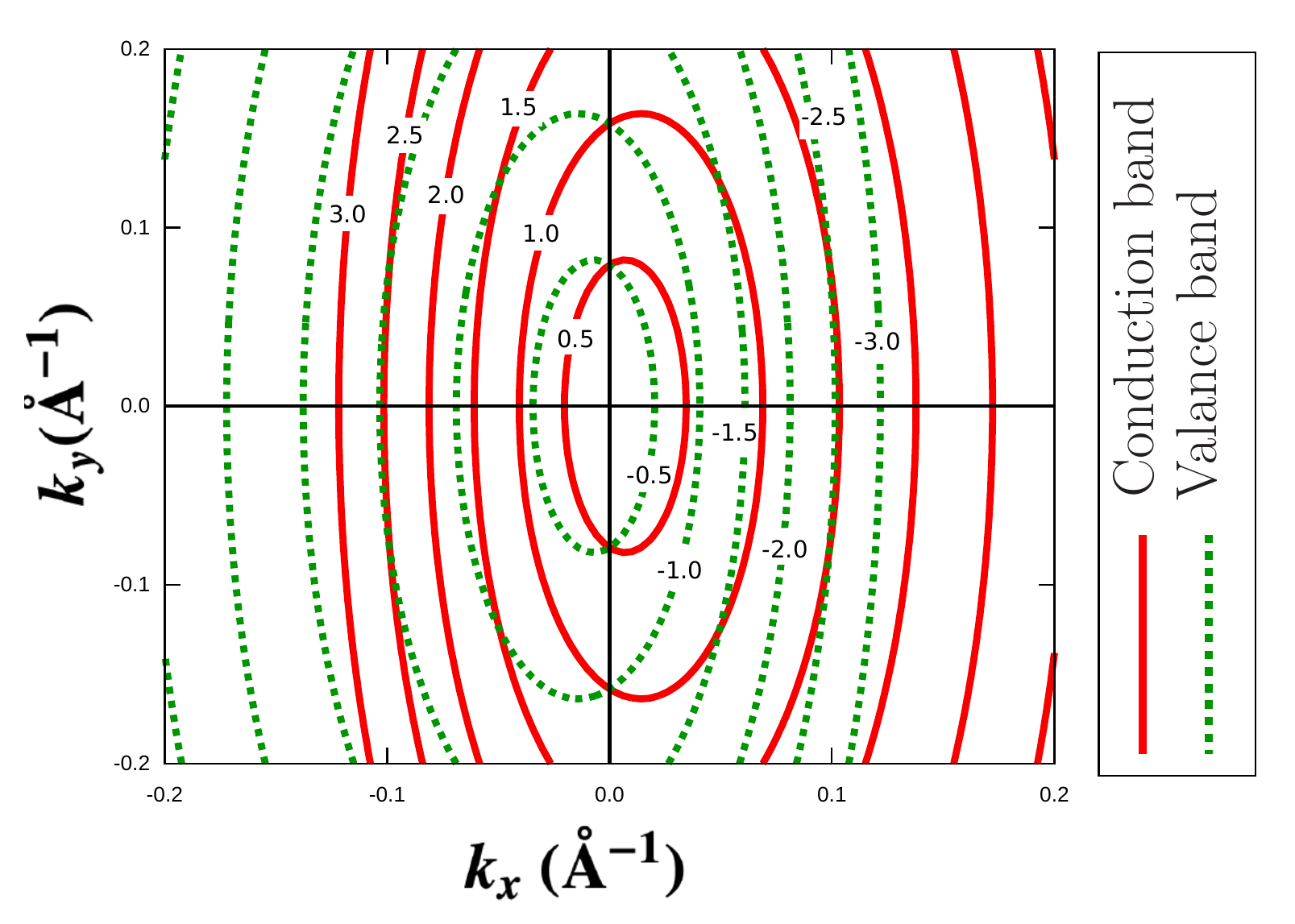}
\caption{Upper panel [(a)] Plot of the anisotropic, tilted energy band dispersion of borophane in the vicinity of ${\bf{k}}_D$ point, in the $k_x$-$k_y$ plane. Lower panel [(b)] Plot showing Fermi surface for conduction band (CB) and valence band (VB) respectively. The Fermi surfaces are tilted along $\pm x$ direction for CB and VB accordingly, with the ratio of semi-minor and semi-major axes being identical for both the bands. Representing values on the contours are in eV.
\label{disp1}}
\end{center}
\end{figure}

%

%


In this section, we study the ballistic transport of Dirac 
electrons across a normal metal-barrier-normal metal (NBN) junction 
in borophane, as shown in Fig.\ref{nbn}. Since borophane possesses lower crystal symmetry, in contrast to graphene, the low-energy effective
Hamiltonian contains asymmetric velocity parameters which results in the
two tilted Dirac cones at the Dirac points ${\bf{k}}_D=(\pm 0.64,0)\text{\normalfont\AA}^{-1}$[\onlinecite{tightb1}--\onlinecite{tightb5}]. The unit 
cell of the borophane contains four atoms and the Bravais lattice constants are $(a_x, a_y)=(1.92, 2.81)\text{\normalfont\AA}$ [\onlinecite{tightb4}] as shown in Fig.\ref{unitcell}.

The low-energy effective
Hamiltonian for normal borophane in the vicinity of the Dirac point is given by [\onlinecite{tightb4}]
\bqa
\mathcal{H}_0=\hbar \sum_{\bf{k}} {\psi_{\bf{k}}}^{\dagger}[v_x k_x \sigma_x+v_y k_y \sigma_y+v_t k_x \sigma_0]\psi_{\bf{ k}}
\eqa 
where ${\psi_{\bf{k}}}={(a_{\bf{k}},b_{\bf{k}})}^{T}$, here $a_{\bf{k}}$ and $b_{\bf{k}}$ stands for the annihiliation
operators of the Bloch states of the two triangular sublattices in borophane. $\sigma_{0,x,y}$ represents usual two dimensional identity and Pauli matrices for the pseudospin space respectively. The Hamiltonian describes an anisotropic $2D$ tilted Dirac cone, specified by $v_x$, $v_y$ and $v_t$, the velocities in the $x$ and $y$ directions and the degree of tilting in the $x$ direction respectively. Typical values for the velocities, in units of  $(\times 10^5 m/s)$, are $v_x=19.58$, $v_y=6.32$ and $v_t=-5.06$. Diagonalization of the Hamiltonian gives the energy spectrum
\bqa
E_{s,{\bf{k}}}&=& \hbar\left[v_t k_x+s\sqrt{(v_x k_x)^2+(v_y k_y)^2}\right]
\eqa
where $s=\pm 1$ denotes the conduction and valence band respectively. The energy dispersion is shown in the three dimensional plot in Fig.\ \ref{disp1}a. 
The corresponding eigenvectors are 
\begin{eqnarray}
\psi_{\vec k} &=&(1, s \exp[ i \beta_{\vec
k}])^{T} /\sqrt{2},\no\\
\tan (\beta_{\vec k}) &=&\frac{v_y k_y}{v_x k_x} = \frac{v_y}{v_x}\tan (\phi_{\vec k})
\label{eigenfn1}
\end{eqnarray}
where $s = {\rm Sgn}(E-\hbar v_t k_x)$ and ${\rm sgn}$ denotes
the signum function. Here the angle $\phi_{\vec k}$ specifies the direction of ${\vec k}$ with respect to $k_x$ axis in the momentum space.
Due to the presence of the tilting term, the Hamiltonian is not symmetric under the parity operation $k_x\rightarrow -k_x$ and the Fermi surface 
is not a circle. For a fixed energy $E_{s,{\bf{k}}}$, the equation determining Fermi surface can be written as
\bqa
&&\frac{(k_x-k_{s,c})^2}{k^2_{s,a}}+\frac{k_y^2}{k^2_{s,b}}=1,\no\eqa
where
\bqa
&&k_{s,c}=\frac{E_{s,{\bf{k}}} |v_t|}{\hbar|A|}, \ \ k_{s,a}=\frac{E_{s,{\bf{k}}} v_x}{\hbar|A|}, \no\\&&
k_{s,b}=\frac{E_{s,{\bf{k}}} v_x}{\hbar v_y\sqrt{|A|}}
,A=v^2_t-v^2_x<0 \ \ .\label{Fermisurf}
\eqa 
Eq.(\ref{Fermisurf}) represents an ellipse with center at $(k_{s,c},0)$, shifted along $x$ direction. Since the center of the ellipse $k_{s,c}\propto E_{s,{\bf{k}}}$,
 the Fermi surface for conduction and that for valence band are shifted in different way. The Fermi surface for conduction (valence) band is a shifted ellipse along the $+(-) x$ direction as shown in Fig.\ \ref{disp1}b. But the ratio of semi-minor and semi-major axes $k_{s,b}/k_{s,a}=\sqrt{(v^2_x-v^2_t)}/v_y\sim 3$ being independent of band index $s$, remains identical for both of the bands and is a characteristic property of the material borophane.

In the presence of barrier, the Hamiltonian can be written as \bqa \mathcal{H}=\mathcal{H}_0+U \eqa where the last term $U$ represents the barrier potential contribution from the external field. Electrons 
with an energy $E=\mu+eV$ where $\mu$ is the chemical 
potential and $V$ is the applied bias voltage are 
incident upon the barrier. 

We shall focus on the transport in the $x$ direction and henceforth assume that the system is translationally invariant along the $y$ direction. The parallel momentum $k_x$ for the electron with incident energy $E$ and conserved transverse momentum $q$ can be determined from the solution of the following relation
\bqa
A k^2_x+B k_x+C=0
\label{kquad}
\eqa
where $B=2 |v_t| E/\hbar$ and $C=(E/\hbar)^2-v^2_y q^2$. We note that since the Eq.(\ref{kquad}) admits two solutions for $k_x$ corresponding to any given energy $E$ and transverse momentum $q$, we have to identify the propagating right moving mode as the incident one. This is achieved by requiring the right (left) moving modes to be finite when $x\rightarrow \infty$ ($x\rightarrow -\infty$) [\onlinecite{obliquekt}]. Substituting $E=E+i \eta$ in Eq.(\ref{kquad}) with $\eta$ being an infinitesimally small positive number, one can identify the right (left) moving mode as the complex root with small $+ve$ ($-ve$) imaginary part. Also the condition for the solution of Eq.(\ref{kquad}) to be real in the normal metal region (Region I: $x<0$, as shown in the Fig.\ref{nbn}) restricts the ratio of incident energy $E$ and transverse momentum $q$ to follow the relation
\bqa
(E/\hbar q)^2>v_0^2; \, \, \, \, v_0^2=|A| v^2_y/v^2_x.
\label{energycond1}
\eqa
To study the transport properties, we shall focus on the electrons in the conduction band in the region I. The wavefunctions in the NBN regions $\psi(x) e^{iqy}$ can be read off from Eq.(\ref {eigenfn1}) and are given by
\bqa
\psi_{I}(x)&=& \frac{1}{\sqrt{2}}
\left(\begin{array}{c} 1 \\
e^{i\alpha_R}\end{array}\right) e^{i k_{R} x}
+\frac{r}{\sqrt{2}} \left(\begin{array}{c} 1 \\
e^{i\alpha_L}\end{array}\right) e^{i k_{L} x}\no\\
\psi_{II}(x)&=& \frac{p_1}{\sqrt{2}}\left(\begin{array}{c} 1 \\
s'e^{i\alpha'_R}\end{array}\right) e^{i k'_R x}
+\frac{p_2}{\sqrt{2}} \left(\begin{array}{c} 1 \\
s'e^{i\alpha'_L}\end{array}\right) e^{i k'_L x}\no\\
\psi_{III}(x)&=& \frac{t}{\sqrt{2}}
\left(\begin{array}{c} 1 \\
e^{i\alpha_R}\end{array}\right) e^{i k_R x}
\eqa
where $k_{R,L}(k'_{R,L})$ are the right and left moving parallel wavevectors in region I (II) and are given by the solution of Eq.(\ref{kquad}) ($E$ replaced by $E-U_0=\delta$), $s'={\rm Sgn}(E-U_0-\hbar v_t k'_{R,L})$, $\alpha_{R(L)}\equiv \alpha_{R(L)}(k_x,k_y)=\beta(k_{R(L)},q)$ 
and $\alpha'_{R,L}({\vec k})=\beta(k'_{R,L},q)$. Here $r$, $p_1$, $p_2$ and $t$ denote the reflection and transmission coefficients at the first and the second interface and can be find out from the boundary conditions of matching the wavefunctions at the interfaces $x=0$ and $x=d$. We have omitted the subscript from $\alpha(\alpha')$ denoting the dependence on the wavevector for brevity and use this notation throughout. One point to mention that the angle of incidence ($\phi_R$) and the angle of reflection ($\phi_L$) for each interface are related to the angles $\alpha(\alpha')$ as evident from Eq.(\ref{eigenfn1}) such as
\bqa
\tan\alpha_{R(L)}=\tan\beta(k_{R(L)},q) &=&\frac{v_y q}{v_x k_{R(L)}} = \frac{v_y}{v_x}\tan (\phi_{R(L)}),\no\\
\tan\alpha'_{R(L)}=\tan\beta(k'_{R(L)},q) &=&\frac{v_y q}{v_x k'_{R(L)}} = \frac{v_y}{v_x}\tan (\phi'_{R(L)}).\no\\
\eqa
The unique asymmetry feature of borophane is reflected in the right and left moving parallel momenta $k_{R,L}$ and accordingly in the angle of reflection ($\phi_L$) and the angle of incidence ($\phi_R$). For graphene with higher symmetry in contrast to borophane, the right and left moving parallel momenta for each interface are related via $k_L=-k_R$ and correspondingly $\alpha_L+\alpha_R=\phi_L+\phi_R=\pi$. This is not valid for borophane which 
indicates that the angle of reflection and the angle of incidence for each interface in case of borophane NBN junction do not follow the rule of mirror reflection. Such unconventional feature is unique to the tilted Hamiltonian. The tilting effect modifies the transmission compared to that of isotropic case which is evident in the detail form of the final transmission coefficient given by
\begin{widetext}
\bqa
t&=& \frac{e^{i (k'_L+k'_R) d}(e^{i \alpha_L}-e^{i\alpha_R})(e^{i \alpha'_L}-e^{i\alpha'_R})e^{-i k_R d}}
{D_t},\no\\
D_t&=&e^{i k'_L d}\left(e^{i (\alpha'_L+\alpha_L)}+e^{i (\alpha'_R+\alpha_R)}\right)
-e^{i k'_R d}\left(e^{i (\alpha'_L+\alpha_R)}+e^{i (\alpha'_R+\alpha_L)}\right)
-s'(e^{i k'_L d}-e^{i k'_R d})\left(e^{i (\alpha'_L+\alpha'_R)}+e^{i (\alpha_L+\alpha_R)}\right)\no\\
\eqa
Transmission probability is computed as 
\bqa
T&=&tt^\ast= \frac{\mathcal{N}}{\mathcal{D}},\, \, \mathcal{N}= 4\sin^2\left(\frac{\alpha'_L-\alpha'_R}{2}\right)\sin^2\left(\frac{\alpha_L-\alpha_R}{2}\right), \no\\
\mathcal{D}&=&\mathcal{N}+2\sin^2\left(\frac{k'_Ld-k'_Rd}{2}\right)
\left[1+\cos(\alpha'_L+\alpha'_R)\cos(\alpha_L+\alpha_R)
+\sin(\alpha'_L+\alpha'_R)\sin(\alpha_L+\alpha_R)
+\cos(\alpha'_L-\alpha'_R)\right.\no\\&&\left.+\cos(\alpha_L-\alpha_R)
-s'\{\cos(\alpha'_R-\alpha_R)+\cos(\alpha'_L-\alpha_L)
+\cos(\alpha'_R-\alpha_L)+\cos(\alpha'_L-\alpha_R)\}\right ].
\label{transprobprop}
\eqa
Eq.(\ref{transprobprop}) reproduces the formula for transmission probability of Dirac electrons in case of graphene based NBN junction 
[\onlinecite{graphenetransmission}]
by imposing the condition on the parallel momenta (in the limit $v_t\rightarrow 0$) such as $k_L(k'_L)\rightarrow -k_R(-k'_R)$. 
\bqa
&&T_{graphene}=
\frac{\cos^2{\alpha_R}\cos^2{\alpha'_R}}{\cos^2{\alpha_R}\cos^2{\alpha'_R}\cos^2{k'_R d}
+\sin^2{k'_R d}(1-s'\sin{\alpha_R}\sin{\alpha'_R})^2}
\eqa
\end{widetext}
In the thin barrier limit $U_0\rightarrow \infty$, $d\rightarrow 0$ with finite value of $U_0 d$, the transmission probability is reduced to
\bqa
T_{\rm{thin}}&=& \frac{1}{\cos^2{\chi}+\gamma^2\sin^2{\chi}}, \,
\gamma=\frac{\sin(\frac{\alpha_L+\alpha_R} {2})}{\sin(\frac{\alpha_L-\alpha_R} {2})}\no\\
\eqa
where $\chi=v_x U_0 d /\hbar |A|$ is the effective barrier strength. 

The ballistic conductance for the system is obtained using Landauer-B\"{u}ttiker's formula [\onlinecite{cond-butt}]
\begin{eqnarray}
G(E) &=& G_0 \int_{-q_{\rm max}}^{q_{\rm max}} \frac{dq}{2
\pi} T(E,q) \label{condx}
\end{eqnarray}
where $G_0=e^2 L_y/\hbar$, $L_y$ being the system size along $y$ direction and $q_{\rm max}=E/(\hbar v_0)$ denotes the maximum 
transverse momenta obeying the relation (\ref{energycond1}). This has been evaluated numerically. The existence of a cut-off in momentum ($q_{\rm max}$) is due to the constraint of incident parallel momentum in normal region to be real. For a fixed energy, $k_x$ in normal region is real for only some of the values of the transverse momentum $q$, as evident from Eq.(\ref{kquad}). This in turn limits the number of momentum modes contributing to the conductance for a particular value of incidence energy.  



\section{Results and Discussion} \label{sec3}
\label{result1}

\begin{figure}[t]
\begin{center}
\includegraphics[width=0.95\linewidth]{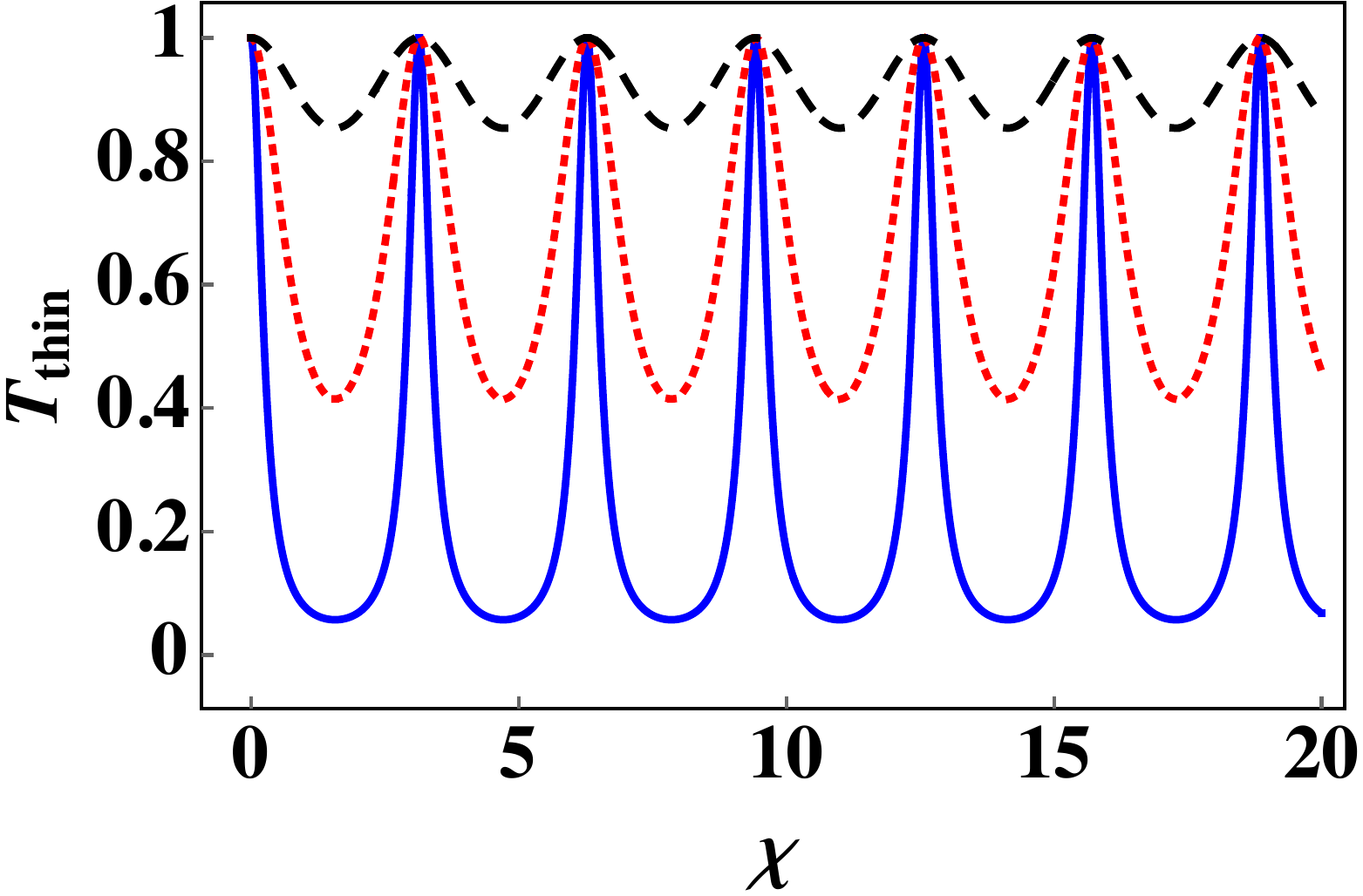}
\caption{ Plot of the transmission probability 
$T_{\rm thin}$ as a function of effective barrier strength $\chi$ for several values of $E$ with
$q=\pi$ (in units of $\text{\normalfont\AA}^{-1}$).
Here $E$(in eV)$=13$ (blue solid line), 
$16.5$ (red dotted line) and 
$33$ (black dashed line). 
The plot shows oscillatory behavior.
\label{fig1}}
\end{center}
\end{figure}

In this section, we shall chart out the results of the corresponding 
theory for transport of Dirac electrons in borophane through a
single barrier along $x$ direction, developed in Sec.\ \ref{form}.
Fig.\ref{fig1} shows the dependence of transmission probability in thin barrier limit on the effective barrier strength for electrons with a finite transverse momentum and different incident energies.
We note that, an electron with higher incident energy crosses the barrier with a greater transmission probability for a specific value of barrier strength, as expected. Also for $\chi\rightarrow n\pi$ ($n=1,2,3,...$), 
the barrier is always transparent ($T_{\rm thin}\rightarrow 1$), irrespective of the incident energy of the electrons. Another interesting point is that the periodicity in $T$ is independent of the incident energy because $E/U_0<<1$ and $U_0$ determines the energy scale of the system.
\begin{figure}[t]
\begin{center}
\includegraphics[width=0.95\linewidth]{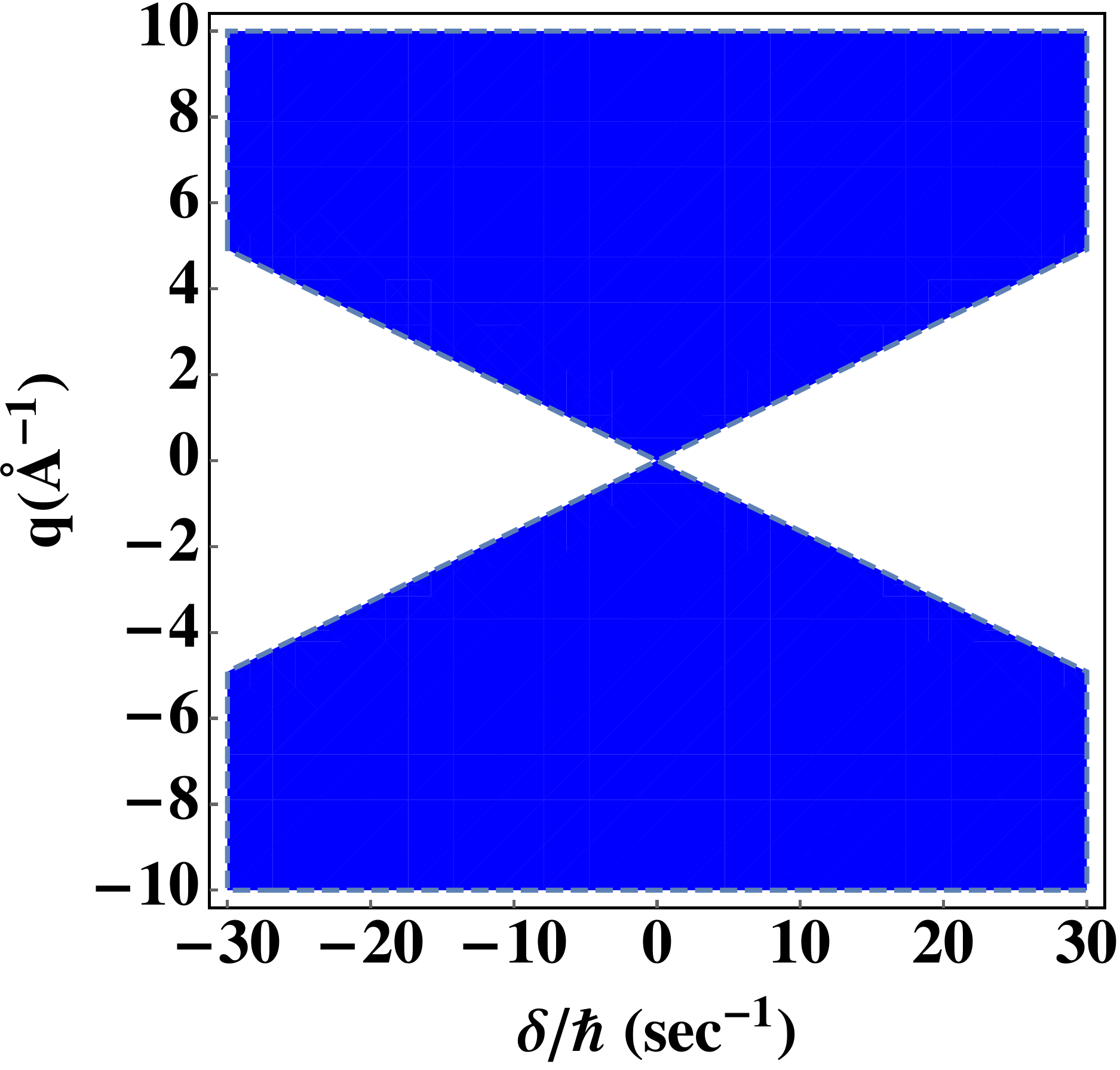}
\caption{"Phase Diagram" showing the plot of the transverse wave vector $q$ as a function of $\delta/\hbar$. The white area is the region of propagating waves (classically allowed region; occurrence of Fabry-Perot resonance) and the 
rest blue area is the region of evanescent waves (classically forbidden; tunneling through evanescent waves). The line of critical momentum $q_c=|\delta|/(\hbar v_0)$ separates the two regions.  
\label{phase}} 
\end{center}
\end{figure}

To discuss the variation of transmission probability as a function of barrier width $d$, we find that the transmission probability may be oscillatory or decaying depending upon the parallel momentum in region II. With real value of $k'_x$, T is oscillatory which is due to the interference effect between the incident and reflected wave in region II. If the interference happens to be constructive, the transmission displays resonances, known as Fabry-Perot resonance. As evident from Eq.(\ref{transprobprop}) that the condition for Fabry-Perot resonance is $(k'_R-k'_L) d=2n\pi$ ($n=0,1,2,3,...$). On the other hand the evanescent waves are present in the barrier region for imaginary value of $k'_x$ and then the transmission probability is exponentially decaying. The decaying behaviour is known as tunneling effect. The condition for appearance of evanescent waves in the barrier region is 
\bqa
(E-U_0)^2-(\hbar v_0 q)^2<0\no
\eqa
Thus the transition between the above mentioned oscillatory and decaying modes can be presented in terms of a critical transverse momentum $q_c= |E-U_0|/(\hbar v_0)=|\delta|/(\hbar v_0)$. The critical momentum is shown in the Fig.\ref{phase}, termed as phase diagram, plotted in the $(\delta/\hbar,q)$ plane. For $|q|<q_c$, the region is oscillatory and on the other hand for $|q|>q_c$, the region contains evanescent waves. The two regions (propagating and evanescent) are separated by the critical momentum line $q=q_c$. The plot being symmetric under $\delta\rightarrow -\delta$, it is evident that for same value of $|\delta|$, the transmission probability must display identical nature irrespective of $E<U_0$ or $E>U_0$. 


\begin{figure}[t]
\begin{center}
\includegraphics[width=0.95\linewidth]{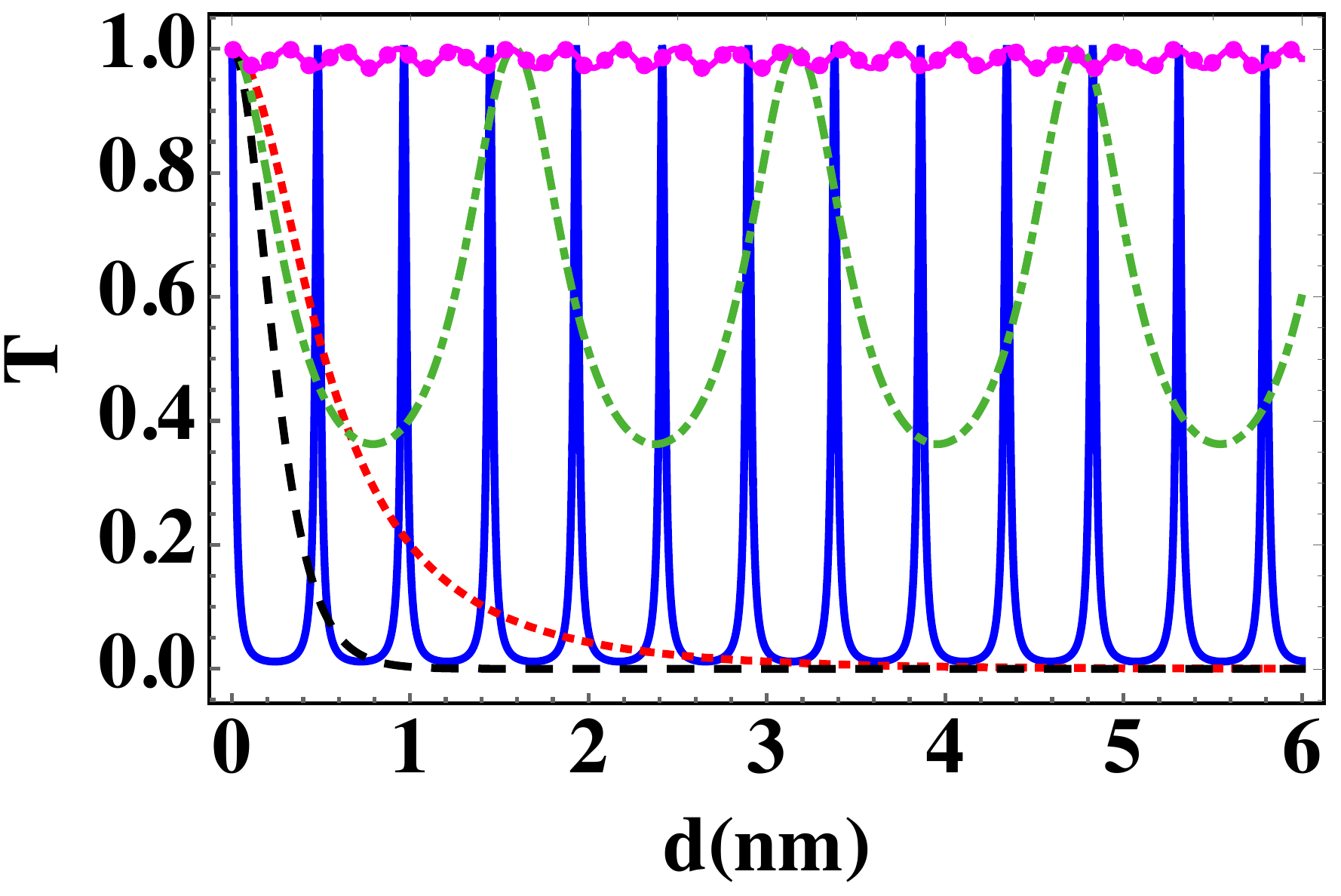}
\caption{ Plot of the transmission probability 
$T$ as a function of barrier width $d$ for several values of $E$ with $U_0=13.2 \,eV$,
$q=\pi/3$ (in units of $\text{\normalfont\AA}^{-1}$).
Here $E$(in eV)$=4.3$ (blue solid line), 
$9$ (red dotted line),  
$13.3$ (black dashed line),  
$18$ (green dotdashed line) 
and $30$ (magenta with large dot line) . 
The plot shows distinct oscillatory and decaying behaviour depending whether $|q|<q_c$ or not. See text for details.
\label{fig2}} 
\end{center}
\end{figure}

In the Fig.\ref{fig2}, transmission probability with different incident energy and fixed values of transverse momenta and barrier strength is plotted against the barrier width. As evident from the phase diagram Fig.\ref{phase}, while $|q|<q_c$, the electrons can propagate through the potential barrier (as shown for $E= 4.3\, eV$ and $E=18\, eV$ in Fig.\ \ref{fig2} and in the other cases, they decay exponentially inside the barrier. For propagating mode, the characteristics of oscillation depends on $E$ for a fixed $U_0$ and $q$, unlike the thin barrier limit. With increase in energy, the period of oscillation increases and amplitude decreases as shown for $4.3$ and $18\, eV$. For $E>>U_0$ (as shown with $E=30\, eV$ case), the amplitude of oscillation decreases significantly and the barrier is nearly transparent irrespective of the barrier width. For the tunneling modes, decay of $T$ depends upon the difference between the incident energy and barrier strength $|\delta|$ as $1/T\sim \sinh^{2}(\frac{v_x d}{4\hbar |A|}\sqrt{(\hbar v_0 q)^2-\delta^2})$. Hence for small value of  $|\delta|$, the decay of $T$ would be sharper with $q$ being fixed, as shown in $E=13.3\, eV$ case. 


One signature of Dirac Fermionic system is reflected in the dependence of the transmission on the critical momentum $q_c$. Due to the dependence of $q_c$ on $|E-U_0|$ for borophane, as pointed out before, it may happen that though $E<U_0$, $T$ is oscillatory (shown by the blue solid line for $E=4.3\, eV$ and $U_0= 13.2\, eV$) and alternatively for $E>U_0$, $T$ is decaying (shown by the black dashed line for $E=13.3\, eV$ and $U_0= 13.2\, eV$). Thus the ratio $E/U_0$ does not play the determining role for transmission to be oscillatory or decaying, in contrast to the case of Schr{\"{o}}dinger quasiparticles.

\begin{figure}[t]
\begin{center}
\includegraphics[width=0.95\linewidth]{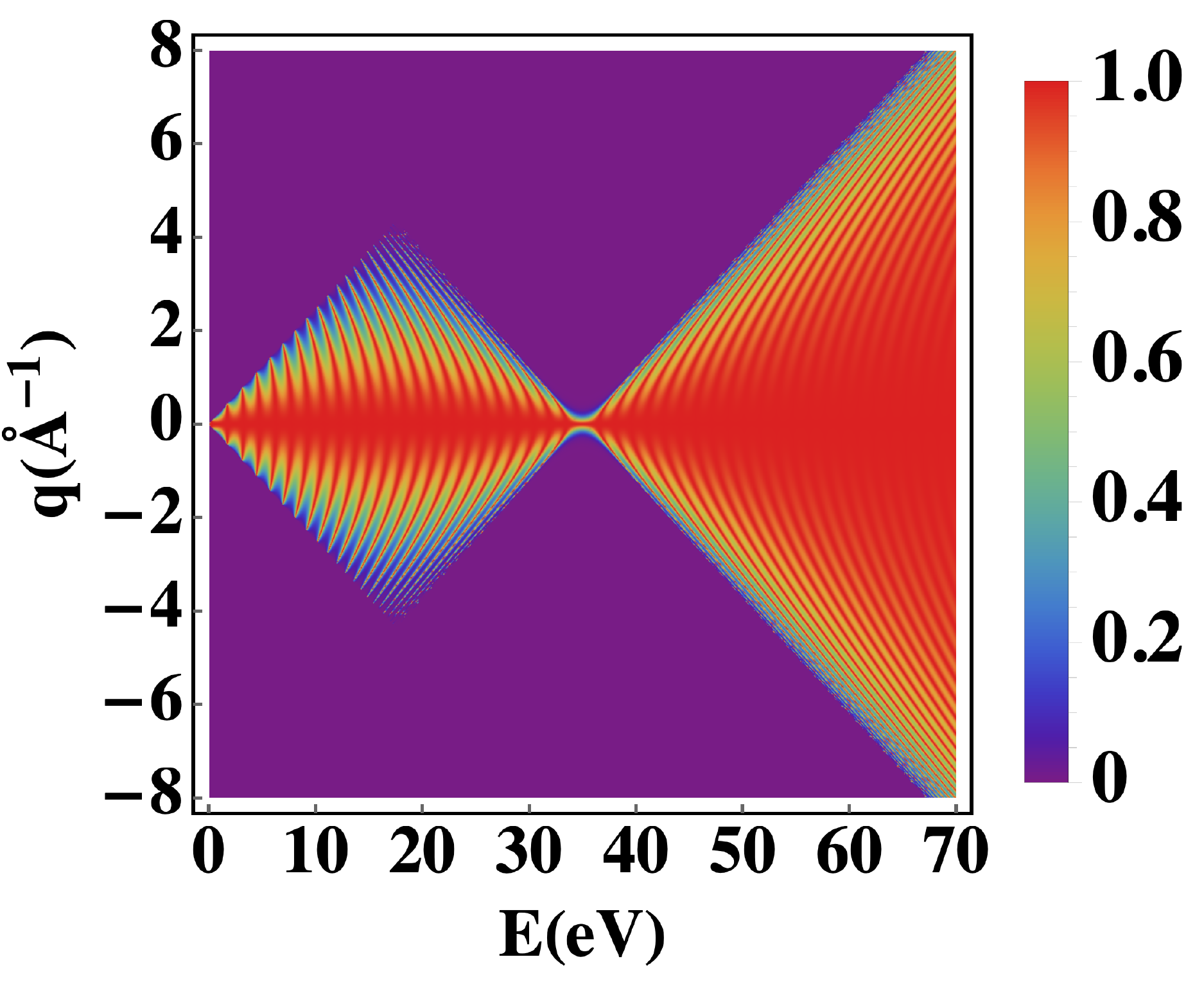}
\caption{Density plot of the transmission probability 
$T$ in $(q,E)$ plane for a fixed value of barrier width $d=2.5 nm$ and barrier strength $U_0=35\, eV$. Incident energy $E$ is plotted in the $x$ axis and transverse momentum $q$ along $y$ axis. The regions of Fabry-Perot resonance, tunneling through evanescent modes are shown. Klein tunneling for normal incidence ($q=0$) is also displayed.
\label{dens}} 
\end{center}
\end{figure}

The presence of evanescent modes leads to a gap in transmission probability determined by the conditions $q^2<q_c^2$. The appearance of the transmission gap is shown in Fig.\ \ref{dens}. In the Fig.\ \ref{dens}, the dependence of transmission probability 
on the incident energy and transverse momentum is studied for a fixed value of the barrier strength and barrier width. Unimpeded penetration at normal incidence $q=0$, known as Klein paradox, is obtained. Then the barrier is fully transparent at any values of incident energy, barrier strength or barrier width. This is the unique signature of any Dirac Fermionic system. As discussed, the existence of the evanescent waves 
in the barrier region 
for transverse momentum satisfying the relation $|q|>q_c$
results in the exponential decay of transmission probability
and in turn forms the transmission gap. As a result, for $E=U_0=35\, eV$, only $q=0$ mode transmits through the barrier.
The regions of Fabry-Perot interference and that of decaying nature are clearly seen in the plot. Another point to note that 
for a particular value of incident energy and barrier strength, the number of momentum modes contributing to the conductance are finite. We mention that, as already discussed, for $E>>U_0$,  the amplitude of oscillation decreases significantly and the barrier is nearly transparent. 

\begin{figure}[t]
\begin{center}
\includegraphics[width=0.95\linewidth]{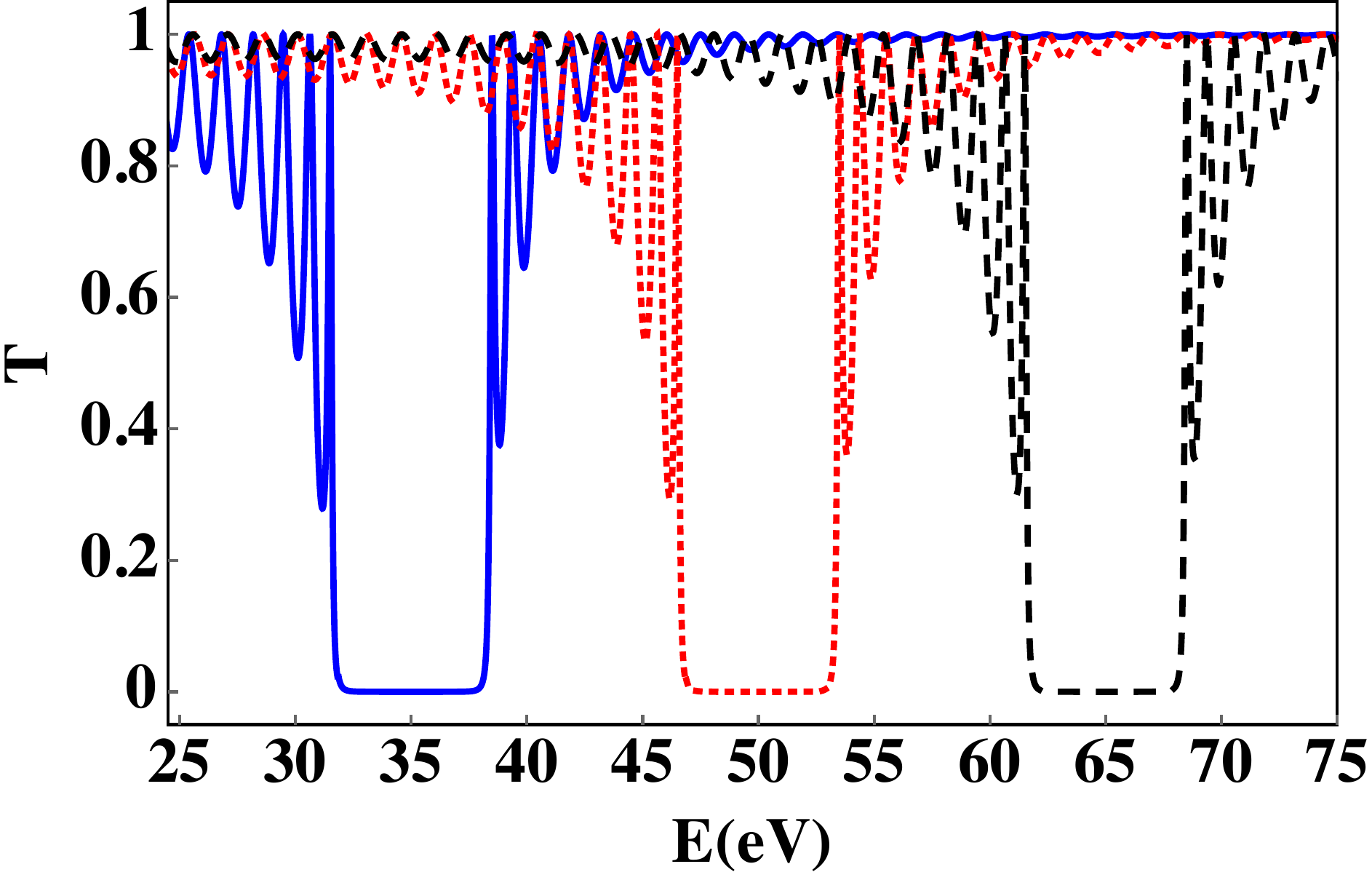}\\
\vspace{1cm}
\includegraphics[width=0.95\linewidth]{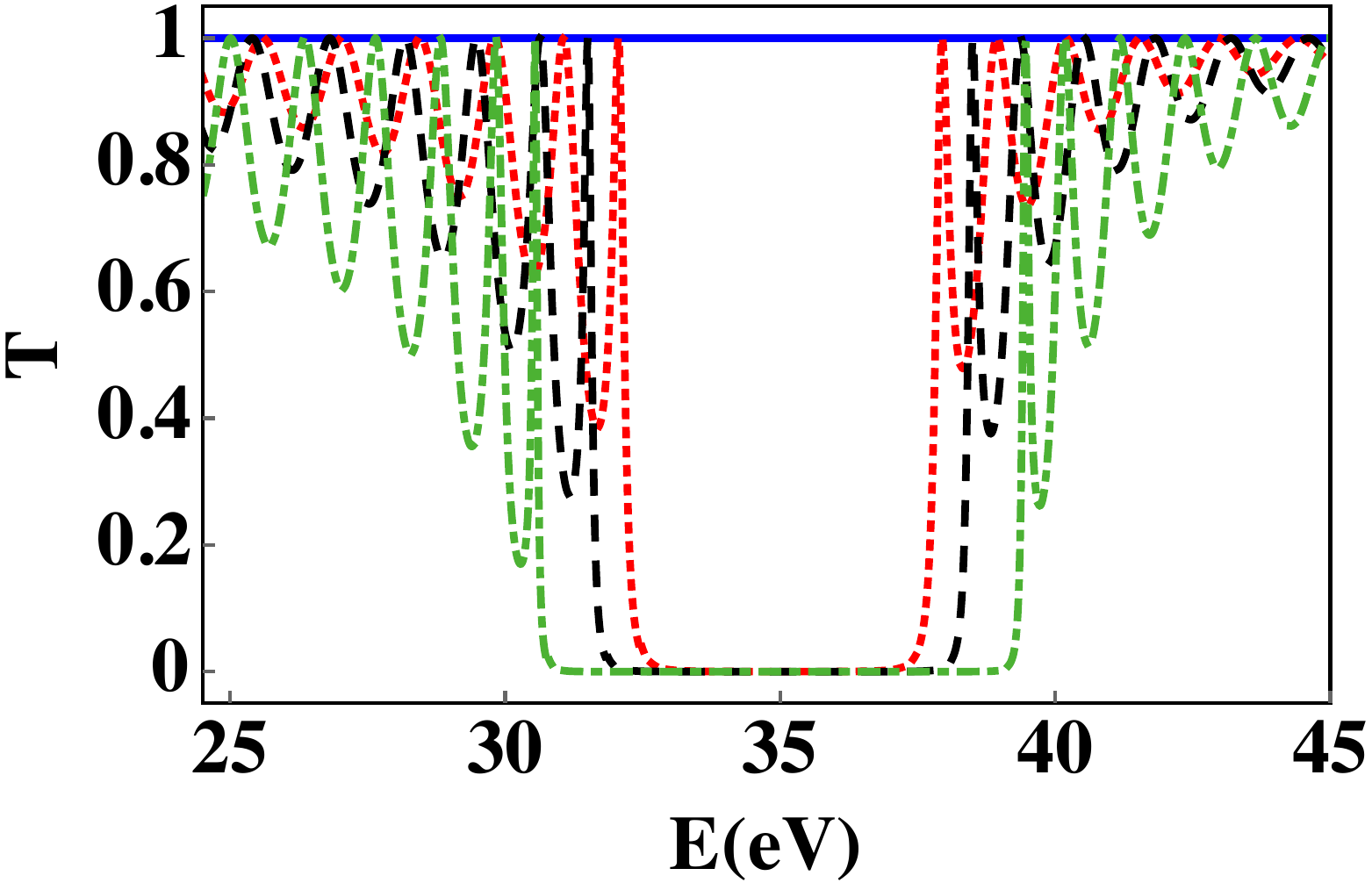}
\caption{ Plot of the transmission probability $T$ as a function of $E$ with $d= 2.5 nm$
for $q=\pi/4 \AA{}^{-1}$ (upper panel) 
and $U_0=35\, eV$ (lower panel). 
In the upper panel [(a)], $U_0=35\, eV$ (blue solid line), 
$U_0=50\, eV $ (red dotted line), 
$U_0=65\, eV $ (black dashed line). 
In the lower panel [$(b)$], 
$q$(in $\AA{}^{-1}$)$=0$ (blue solid line),
$\pi/5$(red dotted line), 
$\pi/4$ (black dashed line), 
$\pi/3$ (green dotdashed line).
We note that transmission gap arises whenever $|q|>q_c$.
\label{fig3}}
\end{center}
\end{figure}

To investigate the dependence of transmission gap on incident energy and transverse momentum, we now study the variation of $T$ as a function of $E$ along constant $q$ (upper panel) and vice-versa (lower panel) in Fig.\ \ref{fig3}. As seen from Fig.\ \ref{dens}, a transmission gap centered at $U_0$ 
occurs for finite values of transverse momentum. The position of the gap shifts along the energy axis as $U_0$ is changed, shown in Fig.\ \ref{fig3}a. Mathematically, the energy gap
can be calculated from the dispersion relation in the barrier region.
Following some straightforward algebra, the gap $\Delta E=2 \hbar v_0 q$. That is why, the width of the gap in Fig.\ \ref{fig3}a remains unchanged for different $U_0$, $q$ being fixed. The variation of the width of the transmission gap with $q$ is depicted in the 
Fig.\ \ref{fig3}b. For normal incidence ($q=0$), Klein tunneling is displayed. However, for a finite value of the transverse momentum, the transmission 
gap sets in and the gap can be tuned by changing $q$.

Since the width of the transmission gap doesn't depend upon the variation of width of the barrier, by tuning the incident energy and transverse momentum of the incident electron, we can successfully control the position and width of the transmission gap, respectively. 
Using this phenomena one can construct a wave vector filter, where only some electrons with specific incident energy are transmitted. 

\begin{figure}[t]
\begin{center}
\includegraphics[width=0.95\linewidth]{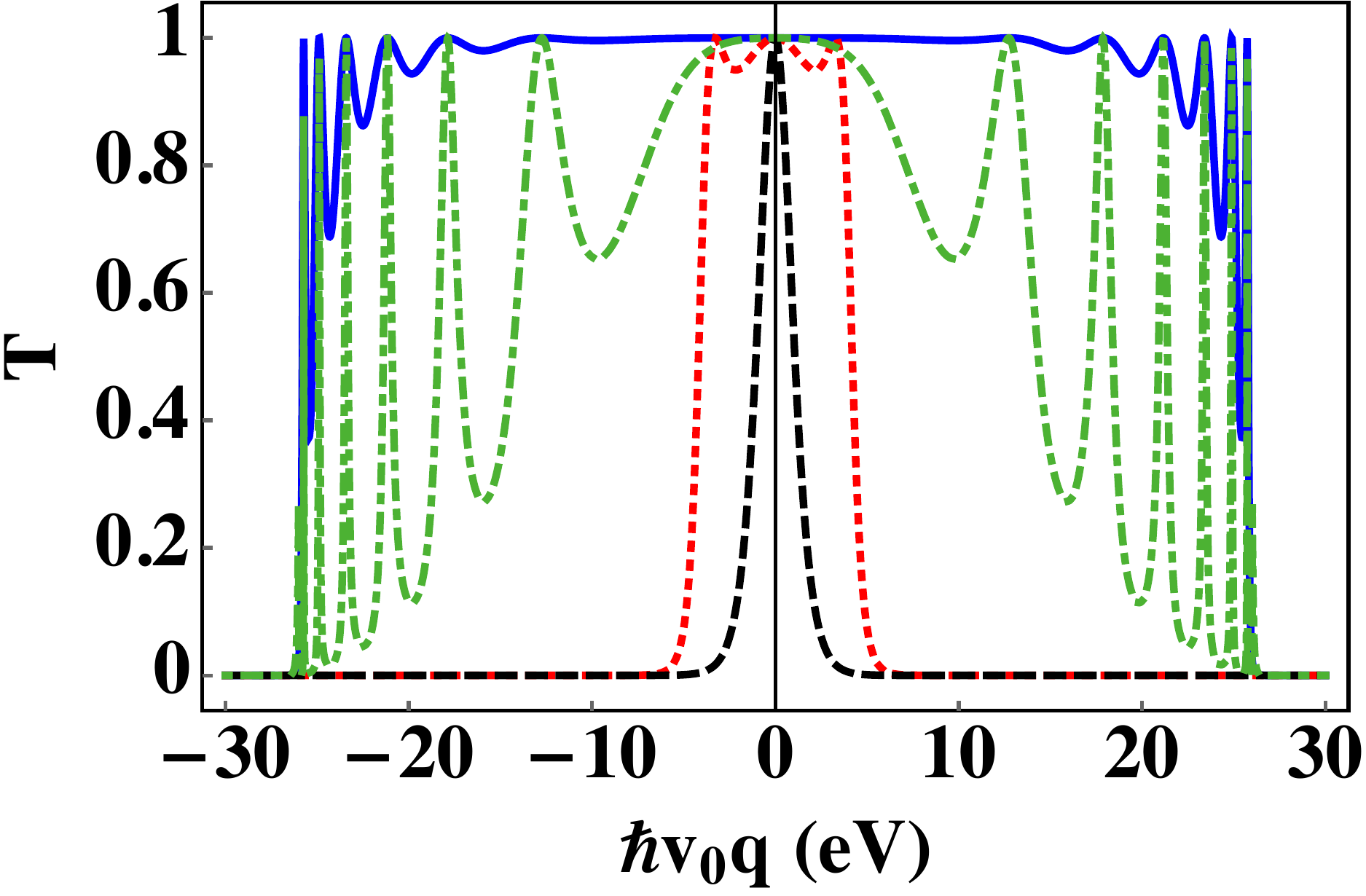}
\caption{Plot of the transmission probability 
$T$ as a function of $\hbar v_0q$ with a fixed value of barrier width $d=1\,nm$ and incident energy $E=31\, eV$ with several values of barrier strength $U_0=5\, eV$ (blue solid line), $U_0=26\, eV$ (red dotted line),
$U_0=31\, eV=E $ (black dashed line) and $U_0=57\, eV$ (green dotdashed line). For a fixed value of incident energy, specific number of modes are transmitted through the barrier.  This indicates the filtering action.
\label{colli}} 
\end{center}
\end{figure}
The filtering effect is explicitly evident when we investigate the variation of the transmission probability as a function of the transverse wave vector for different values of barrier strength as in Fig.\ \ref{colli}. We recognise that the modes with specific value of transverse wave vector are transmitted for a particular value of incident energy and barrier strength. Again, with identical $|\delta|$ ($U_0=5\, eV$ and $57\, eV$), the number of modes are same. The number is smallest for $\delta=0$. We note that the number of transmitted (propagating) modes is quite independent of the barrier width. This enhances the understanding of the filtering action of borophane.

\begin{figure}[t]
\begin{center}
\includegraphics[width=0.95\linewidth]{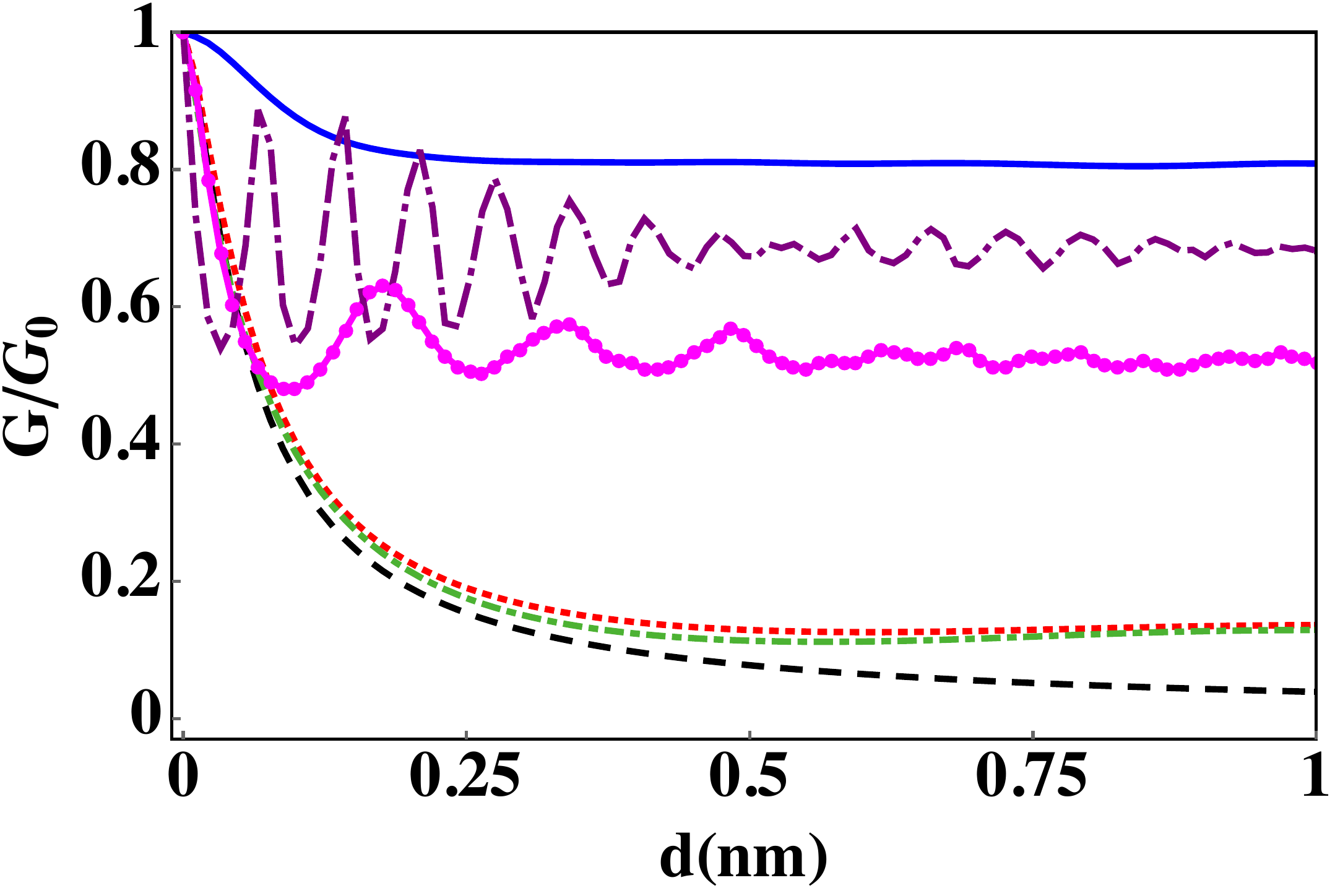}\\
\caption{Plot of the ballistic conductance $G/G_0$ as a function of $d$ with $E= 31\,eV$.
Here $U_0=5\, eV$ (blue solid line), $U_0=26\, eV$ (red dotted line),
$U_0=31\, eV=E$ (black dashed line), $U_0=36\, eV$ (green dotdashed line),
$U_0=57\, eV $ (magenta with large dot line) and $U_0=90\, eV $ (purple double dashed dot line). 
We note that oscillatory or decaying nature in the conductance appears depending upon $E$ and $U_0$. 
See text for details.
\label{fig4}}
\end{center}
\end{figure}
\begin{figure}[t]
\begin{center}
\includegraphics[width=0.95\linewidth]{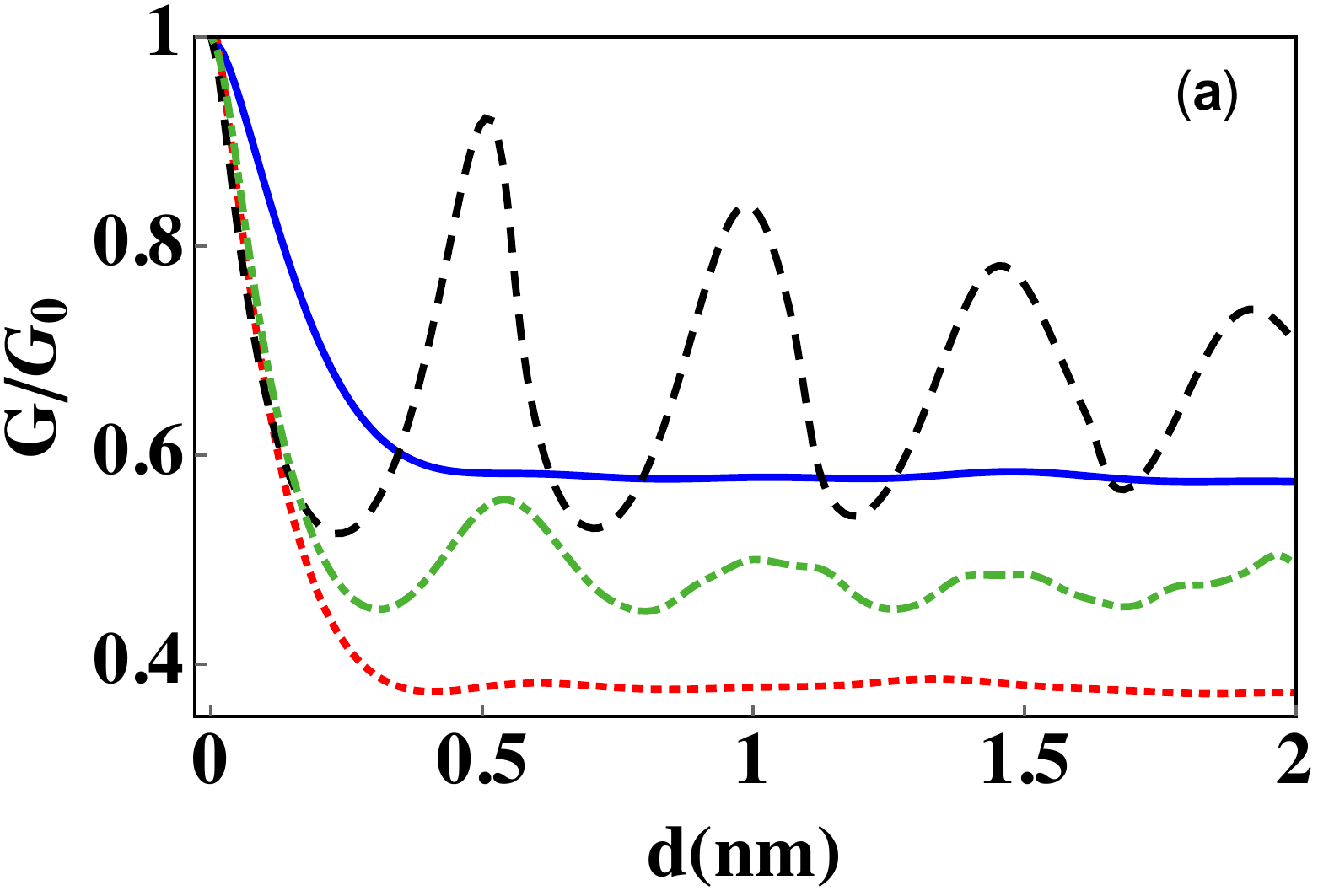}\\
\caption{ Plot of the ballistic conductance $G/G_0$ as a function of $d$ with $|\delta|=|E-U_0|=8.5\, eV$. Different graphs correspond to
$(E,U_0)$(both in eV)$=(13.5,5)$ (blue solid line), 
$(20,11.5)$ (red dotted line), 
$(5,13.5)$ (black dashed line) and
$(11.5,20)$ (green dotdashed line). 
We note that oscillatory or decaying nature appears depending upon $E$ and $U_0$. 
See text for details. 
\label{fig5}}
\end{center}
\end{figure}

To study the nature of ballistic conductance, we plot the conductance against the width of the barrier and incident energy. We want to emphasise that while averaging over transverse momentum in computing ballistic conductance, one have to impose a limit on $q$. This in turn restricts the number of momentum modes contributing in the conductance. The incident energy via $q_{\rm max}$ determines this number of modes. Interplay between the incident energy and barrier strength controls the nature of contributing modes. 
The number of propagating (oscillatory) and evanescent (tunneling) modes ($n_p, n_e$) occupies a significant role to determine the nature of the conductance. The cut-off in momentum mode $q_{\rm max}$ determines the total number of modes contributing to the conductance and thus the total number $(n_p+n_e)$ is proportional to $2q_{\rm max}=2E/(\hbar v_0)$. On the other hand, the number of propagating modes $n_p$ is proportional to $2q_c=2|E-U_0|/(\hbar v_0)$
because whenever $-q_c\le q\le q_c$, the modes are oscillating. Thus depending upon the values of $E$ and $U_0$ ($n_p$ and $n_e$ change accordingly), two types of phenomena determine the nature of the conductance. We can identify two zones depending upon the value of the ratio $\delta/E=(1-U_0/E)$, which is also evident in the phase diagram Fig.\ref{phase}. For smaller value of $|\delta|/E$, the propagating modes are few in number and alternatively with increase in $|\delta|/E$, the number of propagating modes starts to increase. Hence in the first zone with $|\delta|/E<<1$, quantum tunneling through evanescent modes is predominant ($n_e$ exceeds $n_p$) and hence the conductance as a function of barrier width should decay and reaches to a minima. The value of the conductance minima is lowest for $\delta=0$ when $n_p$ is smallest due to filtering effect. On the other hand, with increase in $|\delta|/E$, $n_e$ decreases and in turn the value of the minima increases. If $|\delta|/E$ is increased further ($n_p$ becomes comparable to $n_e$), Fabry-Perot resonance begins to dominate. While discussing the behaviour in the second zone with large value of $|\delta|/E$, one can divide the zone into two sub-zones depending on the signature of $\delta$. Formerly in the first sub-zone with large $+ve$ values of $\delta$, $E>>U_0$ and hence the oscillation in transmission probability becomes very weak and the barrier is mostly transparent. This results in the decay of the conductance but the value of the minima being larger than the previous one with smaller $|\delta|/E$. In the second sub-zone with $\delta/E<<0$, $E<<U_0$, the oscillation in transmission probability is prominent and hence the conductance displays an oscillatory behaviour. In this case, it may happen that transmission occurs through propagating modes only ($n_e=0$) and then the oscillation in the conductance is more pronounced. 

We have studied such situations in two ways- formerly keeping the number of contributing modes $(n_p+n_e)$ fixed (tuning $\delta$) and then with a fixed $n_p$ (changing $E$ and $U_0$, keeping $|\delta|$ fixed). At a fixed $E$, $(n_p+n_e)$ gets determined and then by varying $U_0$, the ratio $\delta/E=n_p/(n_p+n_e)$ can be changed. Since $E$ is kept same in this case, it is sufficient to check the value of $\delta$ to determine the nature of the conductance. For $E=31\, eV$, $U_0=5\, eV$ ($\delta=26$; first sub-zone), the barrier region is transmitted with a high probability and as a result the conductance decays and the minima of the conductance is fairly high, shown in Fig.\ \ref{fig4}. With decrease in $|\delta|$, the minima of the conductance decreases (as shown for $U_0=26 \, eV$ and $\, 36 eV$) and reaches the lowest value for $\delta=0$ i.e; $E=U_0=31\, eV$.  The conductance for $U_0=26 \, eV$ and $\, 36 eV$ coincide because in both cases $|\delta|=5$, as indicated in the phase diagram also. With further decrease in $\delta$, when $U_0=57\, eV$, the conductance is oscillatory as discussed (second sub-zone). The oscillation becomes more prominent by setting $n_e=0$ with $U_0=90\, eV$. 

In Fig.\ \ref{fig5}, $|E-U_0|=|\delta|$ is fixed so that $n_p$ remains unchanged while $(n_p+n_e)$ varies. In this case, we need to check the ratio $\delta/E$ to determine the nature of the conductance. Here for $(13.5,5)$ [$(E,U_0)$ in $\,eV$], since $\delta/E= 0.6$ and hence the conductance should decay being in the first sub-zone. With $(20,11.5)$[$\delta/E=0.4$], the decay of conductance is sharper as a result of the suppression in propagating modes. Large oscillation appears for $(5,13.5)$ because $n_e=0$ as expected, while for $(11.5,20)$, the amplitude of oscillation in conductance is relatively low.


\begin{figure}[t]
\begin{center}
\includegraphics[width=0.95\linewidth]{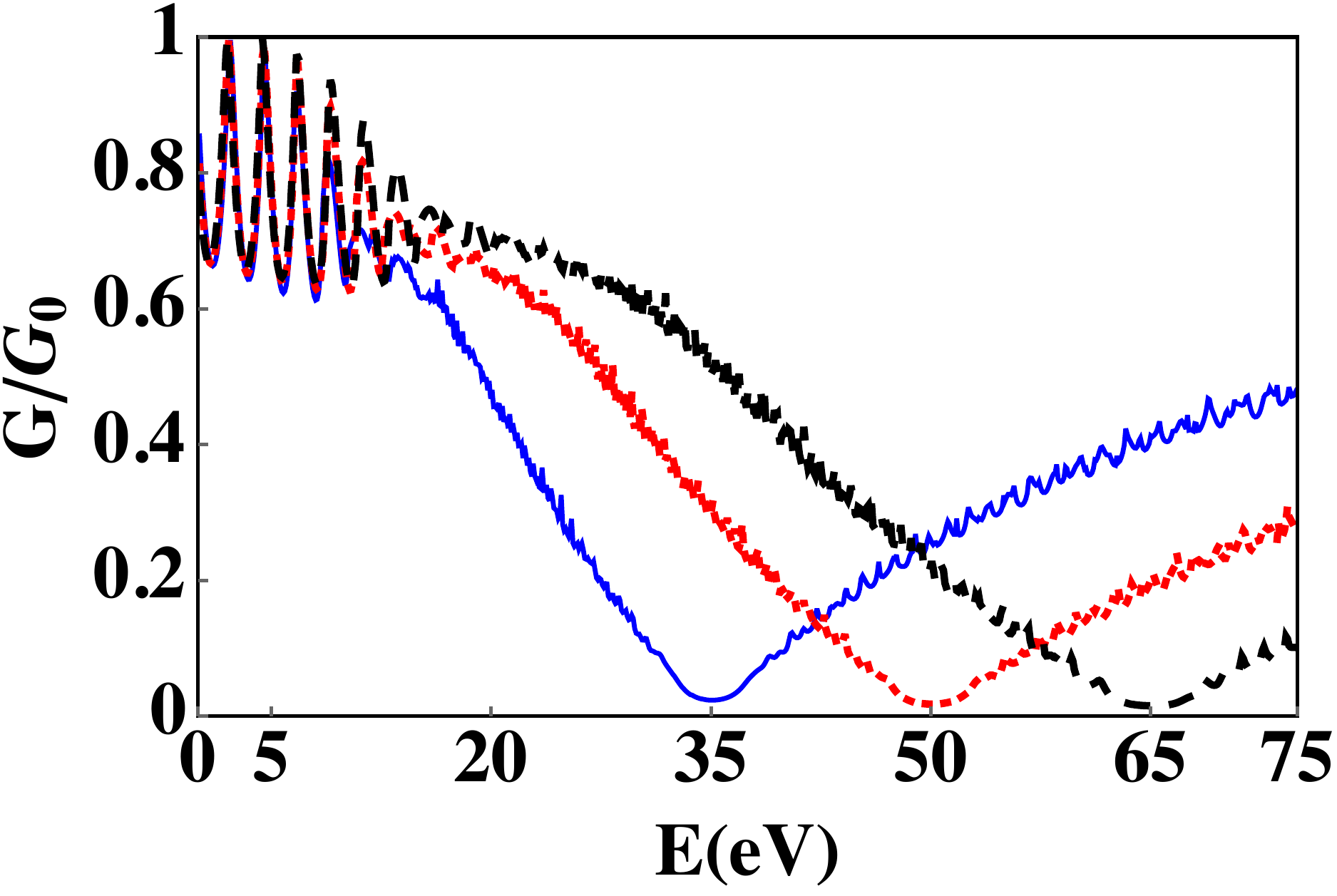}
\caption{ Plot of the ballistic conductance $G/G_0$ as a function of $E$ with $d= 1.5 nm$ for $U_0=35\, eV$ (blue solid line), 
$U_0=50\, eV $ (red dotted line), 
$U_0=65\, eV $ (black dashed line). 
We note that conductance drops to a minimum as a result of transmission gap at $E=U_0$.
\label{fig6}}
\end{center}
\end{figure}
The variation of conductance with the incident energy keeping fixed barrier width for different values of $U_0$ is shown in Fig. \ref{fig6}. We know that a transmission gap centered at $E=U_0$ appears in transmission spectrum as a result of existence of evanescent modes.  This causes a suppression in the conductance to a minimum value for $E=U_0$ and 
position of the minima shifts along the energy axis as $U_0$ is varied. One point to mention that resonance modes in the conductance are present. For $E\le U_0/2$, only propagating modes are contributing to the conductance ($n_e=0$) and hence the conductance displays a strong resonance in this region. We note that by changing $U_0$, one can also modulate the width of the strong resonance region in conductance. The period of oscillation in conductance depends upon the barrier width, as expected. Besides, for obvious reasons, the minima of the conductance decreases with the increase of barrier width. Similar feature can be obtained with variation in barrier strength also.
Hence we can modulate ballistic conductance by tuning incident energy and barrier strength.

\section{Conclusion} \label{sec4}
\label{sec5}
In summary, we have investigated the transmission probability and ballistic tunneling conductance of a single layer borophane NBN junction. The tilted nature of borophane, in contrast to graphene, is expressed through the broken mirror symmetry by angle of incidence and angle of reflection and by the right and left moving parallel momenta in all the regions of NBN junction. We have identified how the tilted term of borophane affects the transmission probability and the conductance. The transmission probability depends on the barrier width, incident energy of the particle, barrier strength and the transverse momentum of the incident particle. We show that there exists a critical value of transverse momentum which in turn determines the nature of transmission probability as a function of barrier width. We present a phase diagram to distinguish the oscillatory and decaying phases in transmission probability in terms of the critical transverse momentum. The transmission probability shows distinguished behaviour as a function of barrier width depending upon whether the transverse momentum is smaller than the critical momentum or not. The analytical expression for the critical momentum has been found. It depends upon the barrier strength, incident energy and the velocities $v_x$, $v_y$ and $v_t$. For $|q|<q_c$, the electrons are in propagating mode and hence the transmission probability becomes oscillatory while in the other case, it decay exponentially inside the barrier. The nature of transmission probability can be predicted from the "phase diagram", displayed in $(\delta,q)$ plane. While observing the variation of transmission probability with incident energy, we note the existence of a transmission gap. 
The position and the width of the gap are theoretically calculated. By tuning the parameters like barrier strength or the transverse momentum of the incident particle, filtered transmission may happen. The filtering effect is explicitly evident when we investigate the variation of the transmission probability with the transverse wave vector. We identify that the modes with specific value of transverse wave vector are filtered for a particular value of incident energy and barrier strength and also the number of transmitted modes is quite independent of the barrier width. This enhances the realisation of the tunable wave vector filtering action of borophane. The filtering effect is also shown in the tunneling conductance spectra. The tunneling conductance drops very rapidly when incidence energy becomes identical to the barrier strength as a result of the suppression in transmission probability due to evanescent modes. This filtering effect is the central result of our work. There exists a lower and upper bound on the possible contributing transverse momentum modes of the incident particle ($\pm q_{max}$) while computing  the conductance. $q_{max}$ is determined by the incident energy and the velocities $v_x$, $v_y$ and $v_t$. The interplay between the Fabry-Perot resonance and the tunneling effect is the determining factor of the nature of conductance as a function of barrier width. This can be expressed through critical momentum and $q_{max}$ . With a fixed incident energy, the total number of modes, contributing to the conductance, gets fixed and then by varying the barrier strength, the number of propagating (oscillatory) and evanescent (tunneling) modes can be controlled and depending upon their ratio the conductance is oscillatory or in decaying regime. The oscillatory features of tunneling conductance are caused by the interference of the Dirac Fermions in barrier region. Our results clearly demonstrate that the tunneling conductance drops very rapidly in the case when the incident energy varies and the barrier width remains fixed in comparison to the situation of 
varying barrier width keeping incident energy fixed. As a consequence, it is easier to get prominent filtering effect by tuning the energies like incident energy or the barrier strength rather than by modulating barrier width. Thus the existence of the transmission gap enables to design a tunable wavevector filter in borophane based electronic devices. Due to very fast progress of experimental technologies, we anticipate to observe such wavevector filtering effect of borophane in the near future.

\section{Acknowledgement}
PD would like to acknowledge Krishnendu Sengupta, Moumita Deb, Tamali Roy for useful discussions. PD thanks funding support from the UGC fellowship, no. 2121551172 (2015), India.

\end{document}